\documentclass[conference,10pt,letterpaper]{IEEEtran}
\usepackage{cite}
\usepackage{amsmath,amssymb,amsfonts}
\usepackage{algorithmic}
\usepackage{graphicx}
\usepackage{textcomp}
\usepackage{xcolor}
\usepackage{lipsum}
\def\BibTeX{{\rm B\kern-.05em{\sc i\kern-.025em b}\kern-.08em
    T\kern-.1667em\lower.7ex\hbox{E}\kern-.125emX}}

\newcommand{\tn}{\textnormal}

\usepackage{hyperref}

\begin{document}
\bstctlcite{IEEEexample:BSTcontrol}

\title{\fontsize{24}{24}\selectfont{Suffused NV-FETs for Reconfigurable Hardware}}
\author{\fontsize{11}{11}\selectfont Alexander J. Edwards \textsuperscript{1}, Son T. Le\textsuperscript{1}, Thomas Broadfoot\textsuperscript{2}, Samuel W. LaGasse\textsuperscript{1}, \\Collin R. Sanborn\textsuperscript{1}, Nicholas A. Blumenschein\textsuperscript{1}, Aubrey T. Hanbicki\textsuperscript{1}, Adam L. Friedman\textsuperscript{1} \\
\fontsize{10}{12}\selectfont \textsuperscript{1}Laboratory for Physical Sciences, College Park, MD, USA, email: ajedwar@lps.umd.edu \textsuperscript{2}TRAP-ICs, Dallas, TX, USA\vspace{-1ex}
}

\maketitle
\noindent \textbf{\textit{Abstract}}---We define a new category of programmable device, the non-volatile (NV) suffused FET, in response to recent advances in photo- 
[1]-[3] 
and e-beam doping 
[4] 
We demonstrate that the writing mechanism of suffused NV FETs largely reduces the required access circuitry compared with other NV switches, and that with available configurations as a closed switch, open switch, NFET, or PFET, the suffused NV FET is ideal for configuration storage in reconfigurable arrays, reducing FET count in one case by 84.5\%. \vspace{-0.5ex}

\section{\textbf{Introduction}}
\label{sec:intro}

This \textit{circuits} paper is inspired by device literature from two fronts. First, a new mechanism of post-fabrication electrostatic channel doping via charge trapping is being developed
[1]-[3]. 
Applied light or electron-beams drastically increase charge mobility to the doping material, enabling programming. In this work, we categorize these non-volatile (NV) FETs as \textit{suffused}, which means ``permeated, as by light''.

Second, NV devices that can be reconfigured to act as a closed switch, open switch, or FET have recently appeared in the literature 
[5]-[7]. 
These devices are appropriately applied for reconfigurable logic, but very little attention is given to the programming circuity, which will be more unwieldy than the authors expect. In this work we categorize these devices as SWITCH+ devices in that their settings include an open or closed switch \textit{plus} a FET configuration or two.

We demonstrate that the marriage of these ideas, operating suffused FETs as SWITCH+ devices in reconfigurable arrays, is the ideal application of both concepts. We also present a brief device-agnostic analysis of NV switch write mechanisms and requisite programming circuitry and demonstrate that suffused SWITCH+ devices reduce total device count for reconfigurable hardware compared with any other configuration storage mechanism.

\section{\textbf{Suffused FETs as SWITCH(+) Devices}}

\begin{figure*}[t]
    \centering
    \includegraphics[width=0.98\linewidth]{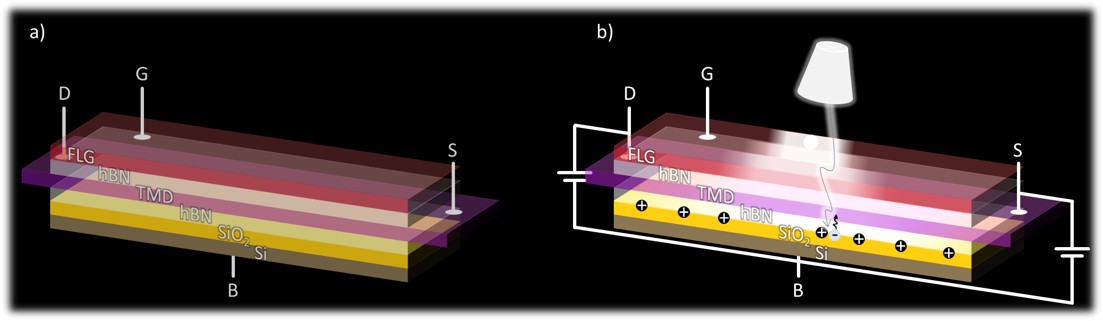}
    \vspace{-1em}
    \caption{A possible four-terminal light-suffused device comprising few-layer graphene (FLG), hexagonal boron nitride (hBN), and a transition metal dichalcogenide (TMD) in an oxidized Si substrate. Normally-insulating hBN traps charge at the hBN/SiO$_\tn{2}$ interface behind a very high WEB. When white light suffuses the device, ionized defects in the hBN create conduction pathways between the hBN/SiO$_\tn{2}$ interface and the TMD channel, enabling charge dopant programming according to $\tn{V}_\tn{SB}$. Removal of the light locks charge at the hBN/SiO$_\tn{2}$ interface, exerting a static $\tn{V}_\tn{dope}$ on the channel.}
    \vspace{-1em}
    \label{fig:1}
\end{figure*}

Suffused FETs have seen recent interest in the literature
[1]-[4], 
and a possible suffused FET construction is illustrated in Fig. \ref{fig:1}. Like floating-gate (FG) or ferro-electric (FE) FETs the suffused FET can be programmed to apply a static electric field on the FET channel, biasing threshold voltage, $\tn{V}_\tn{th}$. However, whereas FG- and FE-FETs are programmed by applying a very large voltage to overcome the high write-energy barrier (WEB) to charge or polarize the doping material, the WEB of the suffused FET is first greatly reduced by suffusing the device with an external energy source (light or electron beams), enabling charge to flow to and from the doping sites according to $\tn{V}_\tn{SB}$. When the suffusion source is disabled, the high WEB is re-established and charge is trapped on the doping material, applying a static doping voltage, $\tn{V}_\tn{dope}$, to the channel.

Using suffusion to control NV FETs can offer several benefits compared with FE- or FG-FETs. A lowered WEB means that charged dopants will be lower-energy compared with those in FE- or FG-FETs, likely increasing lifespan and reliability. Inversely, raising the WEB after programming can increase the charge capacity of the doping material, thereby increasing the memory window. Finally, as we will show in the following subsections, using suffusion to activate devices enables reduction and sharing of programming hardware, \textit{reducing bit cell size to two (in some cases only one) FET}.

\subsection{SWITCH Devices}

Whereas conventional FPGAs use volatile RAM to store the configuration, some systems (e.g. Flash-based FPGAs 
[8]) 
store configuration bits in the states of static switches. We call NV devices that can be programmed to act like open or closed switches SWITCH devices (Fig. \ref{fig:2}a), and they have been proposed using both FGs and FEs
[8]-[11]. 
All NV reconfigurable devices need to operate in two phases: programming and computing. During programming, the configuration of the device is set, and this configuration must not change during computing. Access circuitry needs to accommodate both phases, and device characteristics (number of terminals, write mechanisms, \textit{etc.}) constrain circuit design.

For instance, the conventional programming circuit for a three-terminal FE- or FG-FET SWITCH is illustrated in Fig. \ref{fig:2}b 
[11]. 
To program, the device requires at least one of the source or drain terminals to be grounded, as indicated by the pink transistors which may be shared by multiple bit cells. A word line (WL) is used to select a row of bit cells that are then programmed simultaneously by applying a write voltage to each bit line (BL).

Programming circuitry for suffused FET SWITCHes is illustrated in Fig. \ref{fig:2}c and d. As with the FG- or FE-FETs, one or both of the source or drain needs to be grounded. Suffusion can be global (applied to all devices simultaneously) or local (applied to some devices via on-chip or spatially-multiplexed external sources). Fig. \ref{fig:2}c illustrates programming circuitry for a globally suffused device. All globally suffused devices must be programmed simultaneously, and write-voltages for the whole array can be pre-stored on DRAM-like capacitors, followed by global suffusion. This technique removes the need for on-chip suffusion sources, but will be more prone to error due to capacitor leakage. Fig. \ref{fig:2}d, on the other hand, illustrates programming circuitry for a locally suffused device with an on-chip source. As the WL and selector line (SL) on the suffusion source index the target device, access transistors for the BL are not necessary, and the bit cell therefore requires at most two FETs.

\subsection{SWITCH+ Devices}

\begin{figure*}[t]
    \centering
    \includegraphics[width=0.98\linewidth]{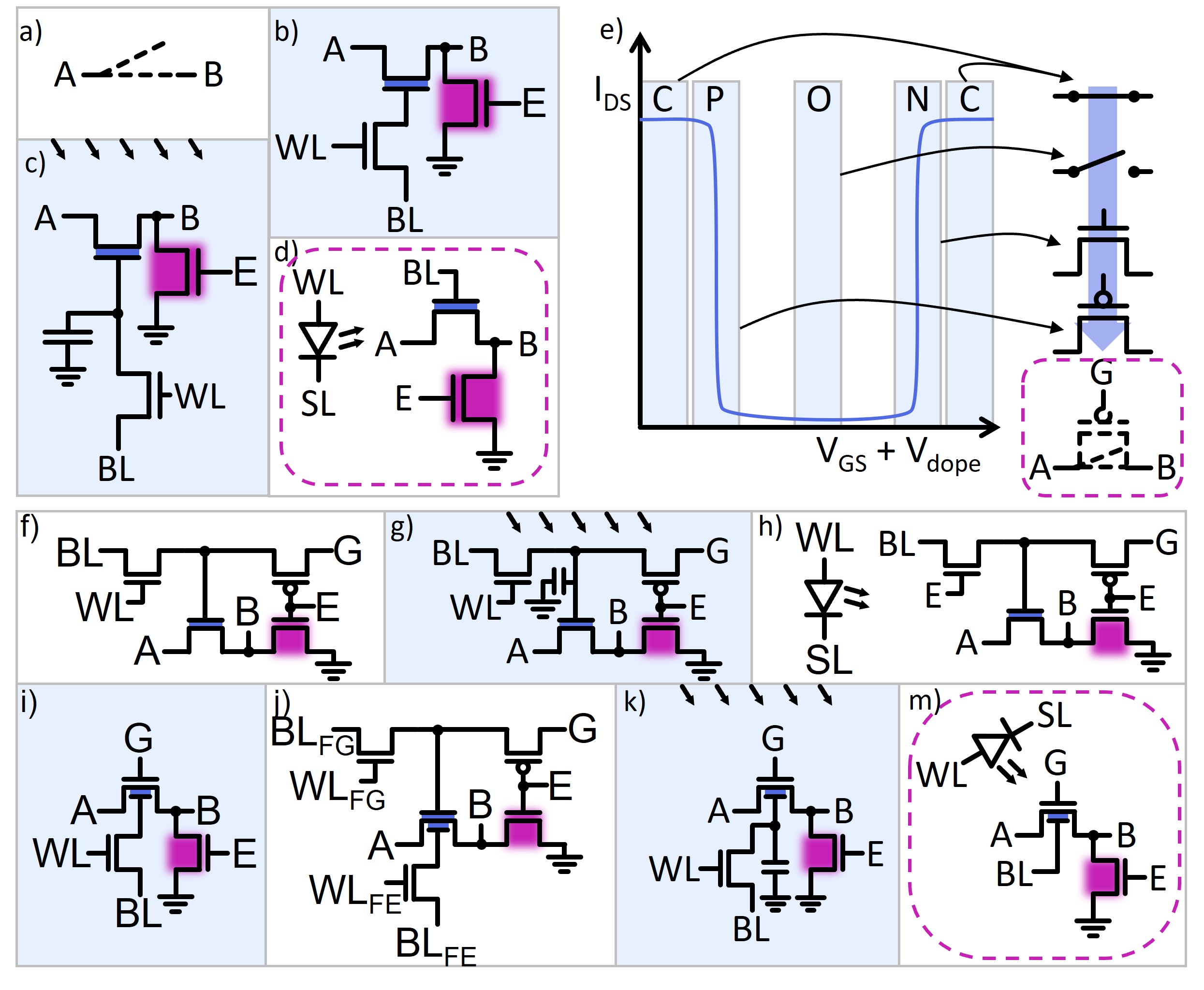}
    \vspace{-1em}
    \caption{a) SWITCH symbol. b-d) SWITCH programming circuitry for a non-suffused device like a FE- or FG-FET (b), a globally suffused FET (c), and a locally suffused FET (d). The locally suffused FET requires the least number of programming transistors. Only 3-terminal devices are shown, as adding another gate does not affect SWITCH behavior, and pink-shaded grounding FETs can be shared by adjacent cells.  e) SWITCH+ IV curve. Modifying $\tn{V}_\tn{dope}$, shifts the operating window to one of four configurations: closed switch, open switch, NFET, or PFET. The SWITCH+ schematic symbol is shown in the bottom right. f-m) SWITCH+ programming circuitry required for each type of device. f-h) circuitry for 3-terminal devices. f) 3-terminal non-suffused FET 
    [6], [7]. 
    g) 3-terminal globally suffused FET. h) 3-terminal locally suffused FET. i-m) Circuitry for 4-terminal devices. i) 4-terminal non-suffused FET. j) 4-terminal FE-FG-FET
    [5]. 
    k) 4-terminal globally suffused FET. m) 4-terminal locally suffused FET, which requires the fewest number of devices.}
    \label{fig:2}
\end{figure*}

We call SWITCH devices that can also be configured as a FET as SWITCH+ devices. Fig. \ref{fig:2}e displays an $\tn{I}_\tn{DS}$ curve of a SWITCH+ device. By changing $\tn{V}_\tn{dope}$, the operational window of the $\tn{I}_\tn{DS}$ curve shifts to embody closed switch (C), open switch (O), PFET (P), and/or NFET (N) behavior. A number of SWITCH+ devices have been recently demonstrated in the literature
[5]-[7], 
though their descriptions of reconfigurable applications and programming circuitry are limited.

To fill this gap, the circuitry required for programming three- and four-terminal SWITCH+ devices is illustrated in Fig. \ref{fig:2}f-m. Because of the dual-purposed gate in three-terminal devices (Fig. \ref{fig:2}f-g), two access transistors are necessary to isolate the compute and programming circuitry, incurring device overhead, making the three-terminal SWITCH+ proposals of
[6], [7] 
less attractive.

Front and back-gates of four-terminal SWITCH+ devices (Fig. \ref{fig:2}i,k,m) are isolated from each other, removing the need for these extra access transistors. Fig. \ref{fig:2}i,j depict circuitry for non-suffused devices, and Fig. \ref{fig:2}k illustrates circuitry for a globally suffused FET. The device requiring the smallest footprint is, however, illustrated in Fig. \ref{fig:2}m, which depicts a locally suffused FET, and as in the previous subsection, selection via the suffusion source enables removal the BL access transistor, and the circuit requires less than two FET-like devices on average to achieve SWITCH+ functionality.

\begin{figure*}[t]
    \centering
    \includegraphics[width=0.98\linewidth]{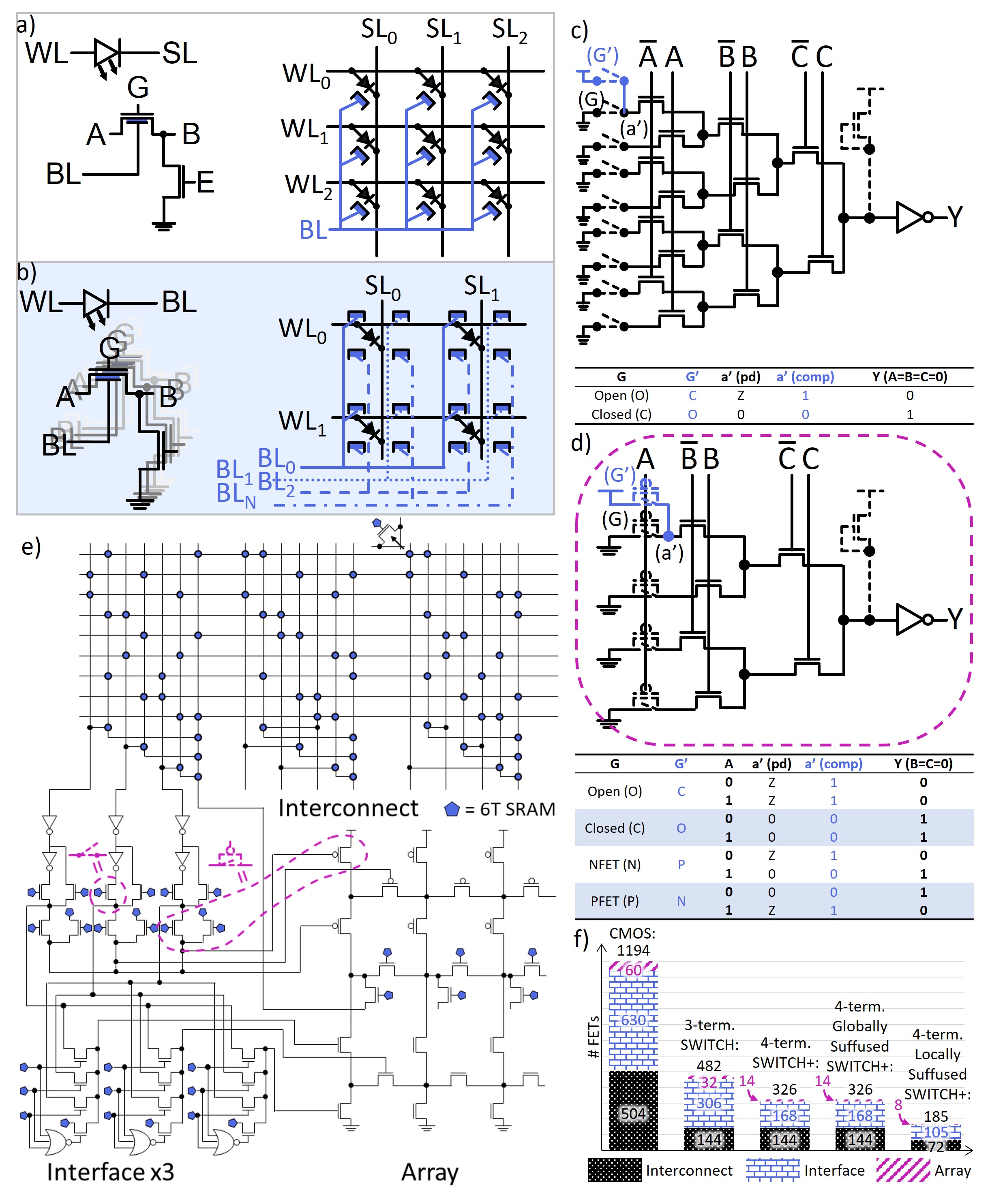}
    \vspace{-1em}
    \caption{a-b) Array circuitry for a 4-terminal locally suffused FET.  Diode-like suffusion sources (\textit{e.g.} LEDs) act as their own crossbar selectors and may be indexed by applying $\tn{V}_\tn{wr}$ to exactly one WL (grounding the rest) and grounding exactly one SL (applying $\tn{V}_\tn{wr}$ to the rest). a) 1x1 array where each source suffuses exactly one bit cell. There is only one global BL. b) 1xN array where each source suffuses N bit cells. In this case there are N BLs, programming all suffused devices simultaneously. c-d) 3-input LUTs comprising SWITCH and SWITCH+ devices, respectively. Both employ a pull-down tree, but faster LUTs can be constructed by adding the circuitry in blue at every branch and removing the dashed pull-up device. No grounding access transistors are necessary since each SWITCH and SWITCH+ device is already connected to a rail.  e) One unit of a TRAP array (modified from 
    [13]). 
    Pink highlights indicate where SWITCH and SWITCH+ substitutions can be made. f) The number of FETs required to construct a TRAP unit with different types of storage device. Using the 4-terminal locally suffused FET reduces transistor count compared with FE- or FG-FET SWITCHes by 61.6\% and compared with CMOS by 84.5\%. }
    \label{fig:3}
\end{figure*}

Fig. \ref{fig:3}a,b illustrates array-level access circuitry for the locally suffused FET.  Fig. \ref{fig:3}a depicts a one-to-one (1x1) addressing scheme where each suffusion source activates exactly one suffused FET. In this case, a global BL can be shared among all the devices. Fig. \ref{fig:3}b depicts a 1xN scheme, wherein a single suffusion source activates the N nearby suffused FETs. N bit-lines are used, one for each device the group, and all N devices are programmed simultaneously. Multiplexing in this way reduces footprint, routing complexity, and latency.

\section{\textbf{Reconfigurable Hardware with SWITCH(+)}}
\label{sec:reconfig}

Most reconfigurable hardware stores configuration bits in SRAM, incurring device overhead. Rather, using NV SWITCH and SWITCH+ devices greatly reduces device count.

\subsection{Lookup Tables}

The most common logic circuit in FPGAs is the lookup table (LUT), and Fig. \ref{fig:3}c,d illustrates a three-input LUT implemented with SWITCH and SWITCH+ devices, respectively. Whereas a three-input CMOS LUT would require 64 FETs, the locally suffused SWITCH version requires 25 FET-like devices in the pull-down configuration or 32 in a faster, more-energy-efficient complementary configuration. In addition, a locally suffused SWITCH+ design further reduces these FET counts to 13 in the pull-down case and 16 in the complementary case. In general, for an N-input LUT, the pull-down locally suffused SWITCH+ network requires only 20\% of the FETs required in the CMOS equivalent circuit.

\subsection{Transistor-level Reconfigurable Hardware}

In addition to conventional reconfigurable logic, proposals for transistor-level (as opposed to gate-level) reprogrammable arrays have recently appeared 
[12]-[14]. 
Fig. \ref{fig:3}e depicts a single unit of a transistor-level programmable (TRAP) array comprising three sections: Interconnect, Interface, and Array. The Interconnect routes signals from other TRAP units to the Interface and is a perfect application for SWITCH devices. The Interface configures the state of each transistor in the Array; to the gate of each NMOS or PMOS in the Array, the Interface drives a constant 0, constant 1, or a logical input from the Interconnect. In other words, the Interface configures whether an Array transistor will behave as a closed switch, open switch, or as a FET, \textit{i.e.} the function of the SWITCH+ device. With careful replacement therefore, \textit{21 CMOS transistors can be replaced by a single four-terminal locally suffused FET}.

\section{\textbf{Case Study: TRAP}}
\label{sec:trap}

For a number of different configuration devices, we calculated how many FET-like devices are required to implement each section of the TRAP unit (Fig. \ref{fig:3}f). Whereas the full CMOS implementation consumes 1194 transistors, a single three-terminal SWITCH device like a FE-FET or FG-FET reduces this count by 59.6\% to 482 FETs. A four-terminal non-suffused SWITCH+ device reduces this further to 326 devices, matched by a four-terminal globally suffused SWITCH+. With the locally suffused FET, this count can be reduced ultimately to 185 FETs for a full TRAP unit, a reduction of 61.6\% compared with a normal FE- or FG-FET SWITCH and 84.5\% compared with CMOS.

\section{\textbf{Conclusions}}

A locally suffused four-terminal SWITCH+ device is the clear winner for next generation NV configuration memory, as its writing method enables elegant device count reductions over every other memory device, including CMOS, FE-FETs, and FG-FETs. In addition, the suffusion writing mechanism removes the memory window/programmability trade-off, enabling arbitrarily high-memory windows, likely to span all four SWITCH+ device functions, without significantly impacting programmability, and retention time and device lifespan can be increased through manipulation of the WEB. The best NV configuration device will therefore likely come from the family of four-terminal locally suffused devices.

\section*{References}

{
\footnotesize
\noindent[1] S. T. Le \textit{et al.}, \textit{2D Mat.} 12.1, 015006, 2024. [2]  M. F. Munoz \textit{et al.}, Proc. SPIE 13593, Phys. Chem. of Semi. Mat. \& Int. XXIV, 1359307, 2025. [3] C. R. Sanborn \textit{et al.}, arXiv:2605.22452, 2026. [4] Q. Lan \textit{et al.}, arXiv:2512.06318, 2025. [5] J. Kang \textit{et al.}, \textit{ACS Nano} 19.13, 12948, 2025. [6] G. Migliato Marega \textit{et al.}, \textit{Nature} 587.7832, 72, 2020. [7] J.-C. Shin \textit{et al.}, \textit{ACS Nano} 20.27, 19439, 2026. [8] Microsemi, \textit{``Design for Performance on Flash-Based FPGAs''}, 2015. [9] Y. Xu \textit{et al.}, \textit{Sci. Ad.} 10.3, eadk1525, 2024. [10] T. Yu \textit{et al.}, \textit{Nat. Comm.} 13.1, 2235, 2022. [11] X. Chen \textit{et al.}, \textit{IEEE TCAS I} 66.5, 1780, 2019. [12] M. M. Shihab \textit{et al.}, Proc. DATE 2019, 528, 2019. [13] A. Fowler \textit{et al.}, \textit{IACR Trans. Crypt. HW \& Emb. Sys.} 2025.2, 579, 2025. [14] C. Sechen \textit{et al.}, US Patent US11362662-B2, 2022.

}

\newpage

\end{document}